\pdfoutput=1
\documentclass[runningheads]{llncs}
\usepackage{url}
\usepackage{amsmath}
\newcommand\UNet{\operatorname{U-Net}}

\usepackage{textcomp}
\usepackage[para,online,flushleft]{threeparttable}

\usepackage{booktabs}
\usepackage{multirow}
\usepackage{graphicx}
\usepackage[table,xcdraw]{xcolor}

\usepackage[normalem]{ulem}
\useunder{\uline}{\ul}{}
\usepackage{subfig}
\usepackage{todonotes}
\usepackage[colorlinks=true,urlcolor= blue]{hyperref}%
\usepackage{bbding}

\begin{document}
\title{Learning shape priors for robust cardiac MR segmentation from multi-view images}
\titlerunning{Robust cardiac MR segmentation with multi-view shape priors}
%
\author{Chen Chen\inst{1}(\Envelope), Carlo Biffi\inst{1}, Giacomo Tarroni\inst{1}, Steffen Petersen\inst{2}, \\ Wenjia Bai\inst{3,4}, Daniel Rueckert\inst{1}}
\authorrunning{Chen. et al.}

\institute{Biomedical Image Analysis Group, Imperial College London, London, UK\\
\and NIHR Barts BRC, Queen Mary University of London, London, UK
\and Data Science Institute, Imperial College London, London, UK
\and Department of Medicine, Imperial College London, London, UK
\\
\email{chen.chen15@imperial.ac.uk}}
\maketitle              
\begin{abstract}
Cardiac MR image segmentation is essential for the morphological and functional analysis of the heart. Inspired by how experienced clinicians assess the cardiac morphology and function across multiple standard views (i.e. long- and short-axis views), we propose a novel approach which learns anatomical shape priors across different 2D standard views and leverages these priors to segment the left ventricular (LV) myocardium from short-axis MR image stacks. The proposed segmentation method has the advantage of being a 2D network but at the same time incorporates spatial context from multiple, complementary views that span a 3D space. Our method achieves accurate and robust segmentation of the myocardium across different short-axis slices (from apex to base), outperforming baseline models (e.g. 2D U-Net, 3D U-Net) while achieving higher data efficiency. Compared to the 2D U-Net, the proposed method reduces the mean Hausdorff distance ($mm$) from 3.24 to 2.49 on the apical slices, from 2.34 to 2.09 on the middle slices and from 3.62 to 2.76 on the basal slices on the test set, when only 10\% of the training data was used.
\end{abstract}
\section{Introduction}
Accurate segmentation of cardiac magnetic resonance (CMR) images is fundamental for assessing cardiac morphology and diagnosing heart conditions \cite{Petersen2017}. Manual segmentation of the anatomical structures is tedious, time-consuming and prone to subjective errors, which is not suitable for large-scale studies such as UK Biobank\footnote{\url{https://imaging.ukbiobank.ac.uk/}} \cite{Bai2018}. Therefore, it is essential to develop automated, fast and accurate CMR segmentation techniques.

Recently, convolutional neural network (CNN) based methods have achieved very good performance for cardiac image segmentation in terms of both speed and accuracy~\cite{Bai2018,Bernard2018,tao2018deep}. However, they may still produce sub-optimal segmentation results in some circumstances. For example, in the Automatic Cardiac Diagnosis Challenge (ACDC)~\cite{Bernard2018}, the top segmentation methods (all CNN-based) achieve high overall segmentation scores for mid-ventricular short-axis slices. However, they sometimes produce poor results or even fail to locate the myocardium in basal slices (due to its more complex shape) and apical slices (due to its small size). This problem is not uncommon and has been reported in the related literature~\cite{zheng20183,Bernard2018,Khened2018}. Methods based on 2D networks, trained in a slice-by-slice fashion, are particularly affected by this problem since they do not incorporate spatial context from neighboring SA images or long-axis (LA) views. On the other hand, 3D networks are capable of incorporating 3D spatial information to perform the segmentation task. Yet the 3D spatial context can be affected by potential inter-slice motion artefacts \cite{tarroni2018comprehensive} and the low through-plane spatial resolution in cardiac SA stacks, thus limiting their segmentation performance. Compared to 2D ones, 3D networks usually contain more parameter and are prone to over-fitting especially when the training set is limited in size since they use 3D volumes rather than 2D slices as input, significantly reducing the number of training samples.

Experienced clinicians are able to assess the cardiac morphology and function from multiple standard views, using both SA and LA images to form an understanding of the cardiac anatomy. Inspired by this, we propose a method which learns the anatomical prior knowledge across four standard views and leverages this to perform segmentation on 2D SA images. The intuition behind our work is that the representation learnt from multiple standard views is beneficial for the segmentation task on the SA slices as different views should share the same representation of the 3D anatomy if they are from the same subject.

The main contributions of this paper are the following:
a) we developed a novel autoencoder architecture (Shape MAE) which learns latent representation of cardiac shapes from multiple standard views;
b) we developed a segmentation network (multi-view U-Net, adapted from \cite{U-Net}), which is capable of incorporating the anatomical shape priors learned from multi-view images to guide the segmentation on SA images; 
c) we assessed the segmentation accuracy and the data efficiency of the proposed segmentation method against common 2D and 3D segmentation baselines by limiting the number of training images, demonstrating that the proposed method is more robust, and less dependent on the size of training data. \\
\\
\noindent\textbf{Related literature.}
A large number of methods have been developed to improve the robustness of the cardiac segmentation. One approach is to learn an ensemble model where the predictions of a 2D and a 3D network are combined~\cite{Isensee2018}. This method is capable of producing accurate results, but has a relatively high computational cost and requires an extra post-processing step to merge the predictions from the two networks. Another approach is to incorporate cardiac anatomical prior knowledge into segmentation networks \cite{oktay2018anatomically,duan2019automatic}. In \cite{oktay2018anatomically}, the learned representation of the 3D cardiac shape is employed to constrain the segmentation model to predict anatomically plausible shapes. The main bottleneck of this method is the requirement of fully annotated 3D high-resolution CMR images which are free from inter-slice motion artefacts and have high through-plane spatial resolution. However, compared to the standard 2D imaging protocol, the 3D one requires the subjects to hold their breath for a relatively long time and therefore is often not feasible for patients with cardiovascular diseases. Instead of using 3D images, we exploit \emph{routinely acquired} 2D standard views to learn the shape representation of the cardiac structures. The learned representation is then injected into a segmentation network to improve its performance on SA CMR images. Of note, the approach in \cite{probabilistic_U-Net} also injects shape priors produced from an autoencoder into a segmentation network. However, the aim of that approach is to generate multiple segmentation hypotheses for ambiguous images, and cannot be readily employed to learn shape priors from different views to enhance cardiac segmentation. 

\section{Methods}
The proposed method consists of two novel architectures: 1) A \textbf{shape-aware multi-view autoencoder} (Shape MAE) which aims at learning anatomical shape priors from standard cardiac acquisition planes incl. short-axis and long-axis views and 2) a \textbf{multi-view U-Net} which performs cardiac short-axis image segmentation by incorporating anatomical priors learned by Shape MAE into a modified U-Net architecture. 
\begin{figure}[htb!]
\centering
\includegraphics[height=0.32\textheight]{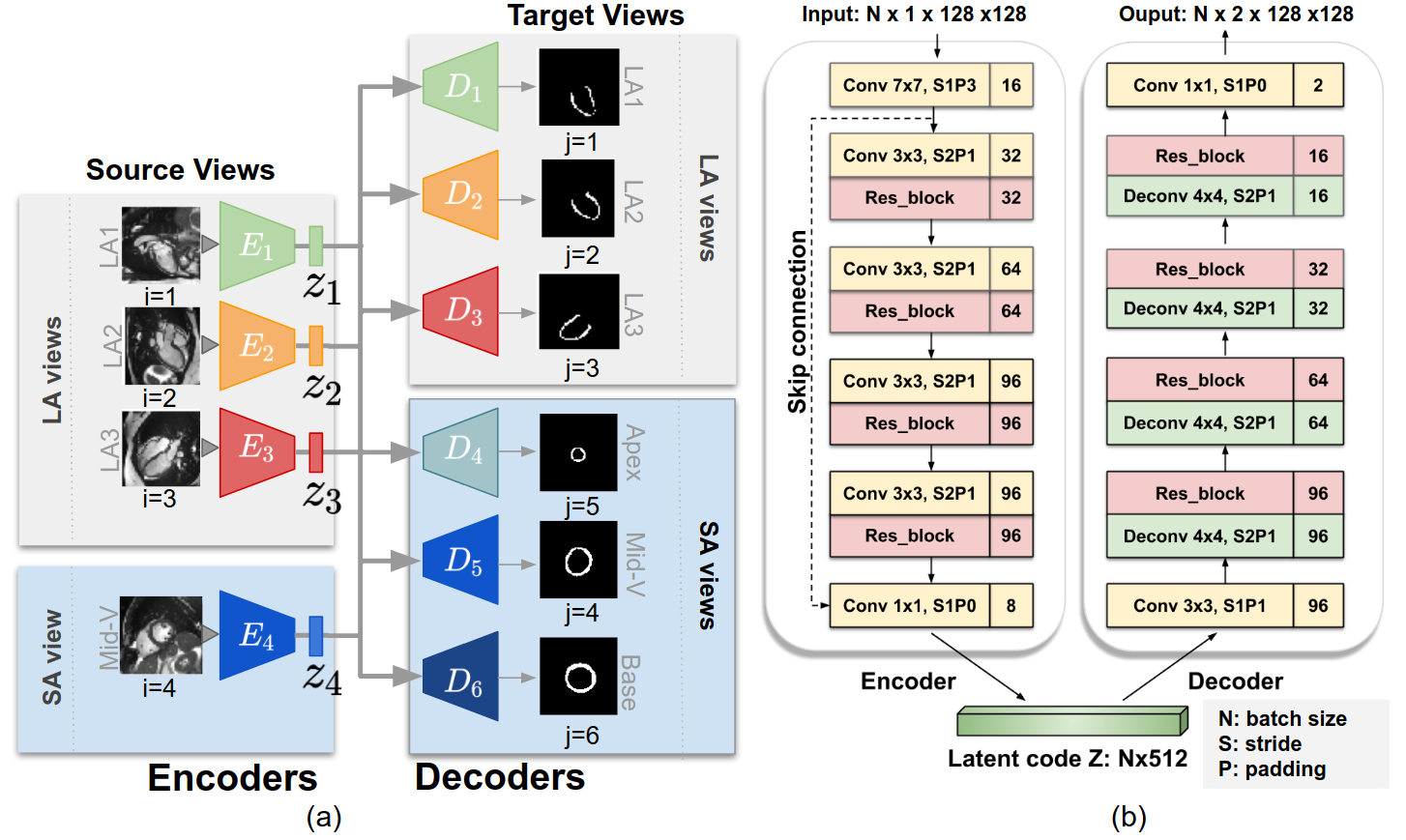}
\caption{\textbf{(a) Overview of Shape MAE. (b) Detailed architectures of each encoder and each decoder}. Each rectangle represents one or a series of convolutional (Conv) or transposed convolutional (Deconv) layers where the number in the square box represents the number of filters for each layer. A `\text{Res\_block}' (pink rectangles) consists of two convolutional layers ($3\times3$) with a residual connection which adds its input to the features from the second layer. Instance normalisation and leaky ReLU activation are applied throughout the network. A sigmoid function is applied to the latent code z to bound its range.}
\label{Fig.MV AE structure}
\end{figure}

\noindent\textbf{Shape MAE: Shape-aware multi-view autoencoder.}
As illustrated in Fig.~\ref{Fig.MV AE structure}, we first present a novel architecture named shape-aware multi-view autoencoder (Shape MAE) which learns anatomical shape priors from standard cardiac views through multi-task learning. Given a source view $X_i$, the network learns the low-dimensional representation $z_i$ of $X_i$ that best reconstructs all the $j$ target views segmentations $Y_j$. In this work, we employ four source views $X_i \; (i=1,\dots, 4)$ which are three LA images - the two-chamber view (LA1),  three-chamber view (LA2), the four-chamber view (LA3) - and one mid-ventricular slice (Mid-V) from the SA view. The target segmentations views $Y_j$ ($j=1,\dots, 6$) correspond to the four previous views plus two SA slices: the apical one and the basal one. All encoders $E_i: z_i=E_i(X_i)$ and all decoders $D_j: Y_j=D_j(z_i)$ in the Shape MAE share the same architecture (see Fig.~\ref{Fig.MV AE structure}~b).

The \textbf{loss function} $\mathcal{L}_\text{Shape~MAE}$ for the whole network is defined as follows:
\begin{equation}
\mathcal{L}_\text{Shape~MAE}= \mathcal{L}_{intra} + \alpha \mathcal{L}_{inter} + \beta \mathcal{L}_{reg}
\label{loss_definition}
\end{equation}

The first two terms of Eq.~\ref{loss_definition} are defined as the cross entropy loss $\mathcal{F}$ 
between the predicted myocardium segmentation $\hat{Y}_{i\rightarrow j}=D_j(E_i(X_i))$ for the target view $j$ given a source image $X_i$ of the same subject and its ground truth segmentation $Y_j$. $\mathcal{L}_{intra}$ denotes the segmentation loss when the source view $X_i$ and the target view $Y_j$ correspond to the same view: $\mathcal{L}_{intra}=\sum_{i=1, i=j}^{4}\mathcal{F}(Y_{j},\hat{Y}_{i\rightarrow j})$, whereas the second term $\mathcal{L}_{inter}$ denotes the loss when two views are different: $\mathcal{L}_{inter}=\sum_{i=1}^{4}\sum_{j=1, i \neq j}^{6} {\mathcal{F}}(Y_{j},\hat{Y}_{i\rightarrow j})$. The third term is a regularisation term on the latent representations $z_i, z_i \in Z$: $\mathcal{L}_{reg}= \frac 1 {|Z|}  \sum_{i=1}^{4}{\left|\left|z_{i} -\bar{z} \right|\right|^2}$, which penalises the L2 distance between $z_i$ and $\bar{z}$, with $\bar{z} = \frac{1}{|Z|}\sum_{i=1}^{4}{z_i}$ being the average $z$ for a subject. Although the latent shape codes from different views of the same subject are not directly shared, this regularisation term forces them to be close to each other.  We use coefficients $\alpha $ and $\beta$ to control the relative importance of $\mathcal{L}_{inter} $ and $\mathcal{L}_{reg}$.
 
The principle behind the proposed network is that different views require \emph{{independent}} functions to map them to the latent space that describes \textbf{{global}} shape characteristics; whereas translating this latent space to another view or plane also requires a \emph{specific} projection function. Predicting the shape of the myocardium based on the six target views instead of a single view encourages the
network to learn and exploit correlations between different views, resulting in
a global, view-invariant shape representation rather than a local representation for a particular
view. All the encoders and the decoders in this framework are trained jointly in a multi-task learning fashion, with the benefit of avoiding over-fitting and encouraging model generalisation~\cite{caruana1997multitask}.\\

\begin{figure}[htb!]
\centering
\includegraphics[height=0.3\textheight, width=1.\textwidth]{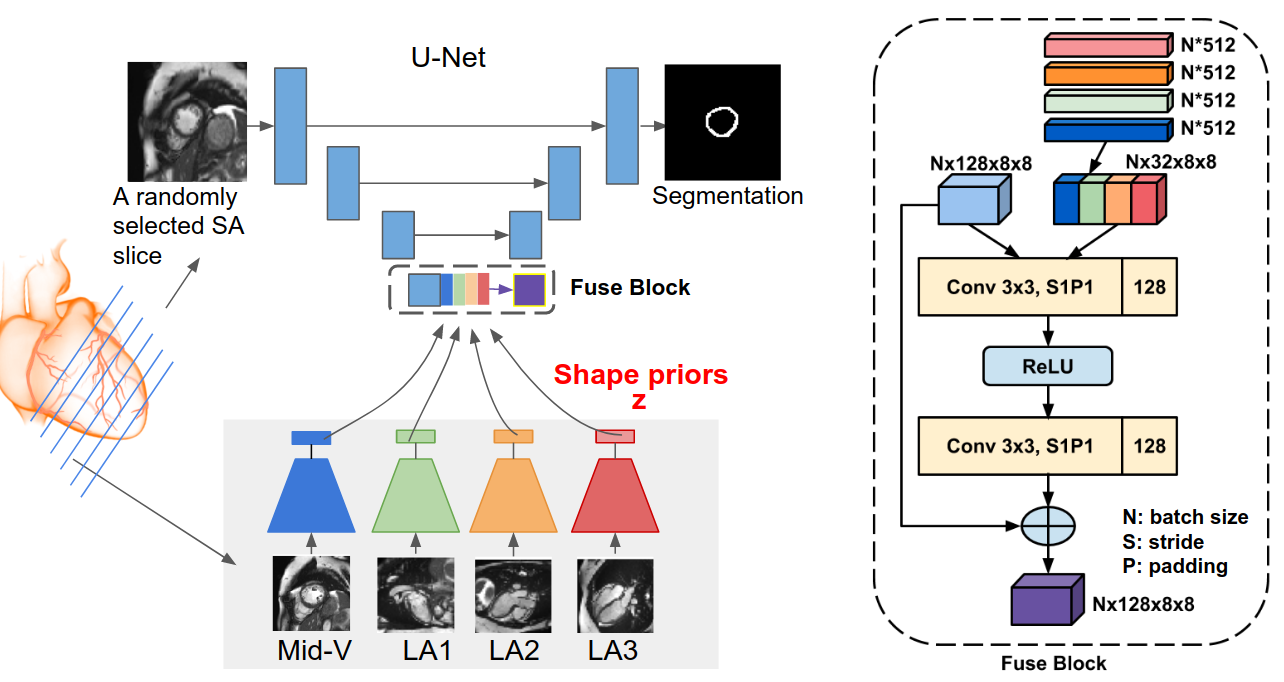}
\caption{\textbf{(a) Overview of the proposed MV U-Net.} \textbf{(b) Architecture of the `Fuse Block'.} The number of shown feature map blocks of the U-Net is reduced for clarity of presentation. Batch normalisation and ReLU activations are applied throughout the network. For each subject, the shape code of each view is reshaped to $1\times8\times8\times8$ and then concatenated with the other three along the second axis to form an input of $1\times32\times8\times8$ to the Fuse Block.}
\label{Fig.MV U-Net}
\end{figure}
\noindent\textbf{MV U-Net: Multi-view U-Net.}
As shown in Fig.~\ref{Fig.MV U-Net}, we propose a segmentation network called multi-view U-Net (MV U-Net) based on the original U-Net~\cite{U-Net} for cardiac SA image segmentation. The proposed network is capable of incorporating the anatomical shape priors learned by Shape MAE. Similar to the original architecture, the proposed architecture comprises 4 down-sampling blocks and 4 up-sampling blocks to learn multi-scale features. Differently from the original U-Net, we reduced the number of filters at each level by four times to account for the fact that cardiac segmentation is simpler than the lesion segmentation (with multiple candidates) which was the task that the original U-Net was applied to. In addition, a module called `Fuse Block' is introduced in the bottleneck of the network (see Fig.~\ref{Fig.MV U-Net}~b) to inject the latent codes into the segmentation network. This fusing approach is different from that in \cite{probabilistic_U-Net} where the latent codes are simply concatenated with U-Net activations. The proposed module consists of two convolutional (Conv) kernels ($3\times3$) and a residual connection to combine the shape representations from different views through learnable weights. Thanks to this module, given an arbitrary short-axis image slice $I^p$ from a subject $p$ and its correspondent shape representations ${z_1^p,z_2^p,z_3^p,z_4^p}$ obtained by Shape MAE (one for each of the four standard views), the network can predict a segmentation $S^{p} = f_{\text{MV}\UNet}(I^p,z_{1}^{p},z_{2}^{p}, z_{3}^{p},z_{4}^{p};\theta)$ by distilling the prior knowledge to the high-level features of the network, allowing it to efficiently refine the segmentations through multi-view information. The network is trained using standard training procedure with a cross entropy loss to optimise the parameters $\theta$ of the MV U-Net. 

\section{Experiments and Results}
\textbf{Cardiac multi-view image dataset.}
Experiments were performed on a dataset acquired from 734 subjects. For each subject, a stack of 2D SA slices and three orthogonal 2D LA images are available. All the LV myocardium were annotated on the SA images as well as the LA images at the end-diastolic (ED) frame using an automated method followed by manual quality control. All the images were acquired using one scanner. The spatial resolution of the images is $1.8 \times 1.8 \times 10~mm$. 

In our experiments, the dataset was randomly split into two subsets: a training set (570 cases), a test set (164 cases). All LA images were registered to a template subject using rigid transformation with MIRTK toolkit\footnote{\url{https://mirtk.github.io/}}. All 2D SA slices have been cropped to the size of $128 \times 128$ pixels where the left ventricle is roughly in the center of every image. Benefiting from the view planning (which is a standard step during the cardiac image acquisition), we simply use the intersection point of the three orthogonal LA images on every SA slice to determine its center of the interest region. All the networks were trained for 200 epochs on an NVIDIA$^{\tiny{\text{\textregistered}}}$
GeForce$^{\tiny{\text{\textregistered}}}$ 2080 Ti, using an Adam optimizer with a batch size of 10. The learning rate for Shape MAE was set to 0.0001 whereas the learning rate for the segmentation network was set to 0.001. In our experiments, $\alpha $ was empirically set to 0.5 and $\beta$ to 0.001 in the $\mathcal{L}_\text{{Shape~MAE}}$. The proposed algorithm was implemented in Pytorch.\\ 

\noindent\textbf{Segmentation results.}
To evaluate the segmentation accuracy, we use two measurements: the Dice score and the Hausdorff distance (HD). The proposed method is compared against: a 2D U-Net~\cite{U-Net}, a state-of-the-art 2D FCN for cardiac MR image segmentation~\cite{Bai2018}, and a 3D U-Net~\cite{Cicek2016}. For fairness and ease of comparison, all models were set with the same number of filters at each level (starting with 16 filters in the first layer) and trained with the same pre-processing and training schedule. For the 3D network, we resampled SA images to a voxel size of $1.8 \times 1.8 \times 1.8~mm$ and cropped each to a size of $128 \times 128 \times 64$ during pre-processing. We trained MV U-Net and the baseline networks with two settings: in one case we used \textbf{10\%} of the training set, while in the other one we used \textbf{100\%}. Of note, in each setting, we first trained a Shape MAE and then trained a MV U-Net where shape priors of four standard views were obtained using corresponding encoders in the Shape MAE.

\begin{figure}[htb!]
    \centering
    \includegraphics[width=\textwidth]{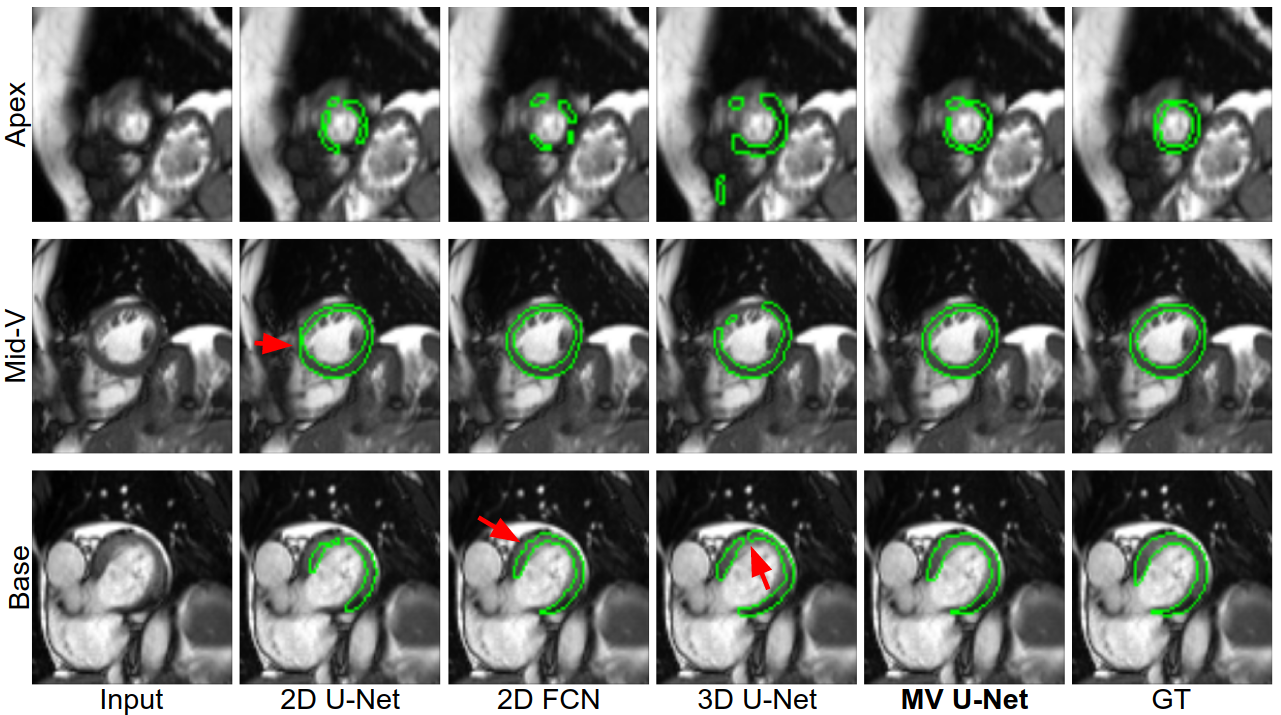}
    \caption{Visualisation of the predicted segmentations and correspondent ground truth (GT) from the baseline models and MV U-Net (all trained with \textbf{10\%} training subjects) on an apical, a mid-ventricular and a basal slice from one patient. Compared to the baseline models, MV U-Net produces more accurate segmentation with stronger spatial coherence.}
    \label{fig:Vis_results}
\end{figure}

Results on the test set are shown in Table~\ref{table: result}. From the table, it can be observed that the proposed method outperforms the baseline models in both the low-data setting and the high-data setting, with improved Dice scores at the apex, middle, and base of the left ventricular myocardium. In particular, when only 10\% data was used, the proposed method reduces the mean HD from 3.24 to 2.49 $mm$ on the apical slices, from 2.34 to 2.09 on the middle slices and from 3.62 to 2.76 on the basal slices, compared to the 2D U-Net. Fig.~\ref{fig:Vis_results} shows examples of the segmentation results from all the networks where the proposed method not only produces more robust segmentation across slices compared to the results from the 2D networks, but also achieves more anatomically plausible results in comparison to the 3D one (see the {\color{red}red arrows} in this figure). Visualization results of the segmentation networks trained in the high-data setting and Shape MAE are provided in the supplementary material.


\begin{table}[!htb]
\centering
\caption{Comparison of the myocardium segmentation accuracy of the baseline models and the proposed method in terms of the mean and the standard deviation of Dice score and HD distance (mm) obtained on the test set (n=164). The comparison has been carried out separately for apical, mid-ventricular, and basal slices.}
\label{table: result}
\resizebox{\textwidth}{!}{%
\begin{threeparttable}
\begin{tabular}{@{}lclllrrr@{}}
\toprule
\multicolumn{1}{c}{\multirow{2}{*}{\textbf{Method}}} & \multicolumn{1}{c}{\multirow{2}{*}{\textbf{\# Training subjects}}} & \multicolumn{3}{c}{\textbf{Dice}} & \multicolumn{3}{c}{\textbf{HD}} \\ 
\multicolumn{1}{c}{} & \multicolumn{1}{c}{} & \multicolumn{1}{c}{Apex} & \multicolumn{1}{c}{Middle} & \multicolumn{1}{c}{Base} & \multicolumn{1}{c}{Apex} & \multicolumn{1}{c}{Middle} & \multicolumn{1}{c}{Base} \\ \midrule

2D U-Net & 57 (10\%) & 0.898 (0.090) & 0.932 (0.035) & 0.923 (0.077) & 3.239 (6.918) & 2.337 (2.913) & 3.617 (9.058) \\
2D FCN & 57 (10\%) & 0.873 (0.113) & 0.926 (0.041) & 0.919 (0.069) & 3.088 (3.882) & 2.317 (1.440) & 2.948 (2.691) \\
3D U-Net & 57 (10\%) & 0.890 (0.083) & 0.923 (0.043) & 0.923 (0.043) & 2.839 (3.980) & 3.573 (9.05) & 4.469 (10.02) \\
MV U-Net & 57 (10\%) & \textbf{0.905 (0.076)} & \textbf{0.932 (0.025)} & \textbf{0.926 (0.088)} & \textbf{2.487 (3.022)} & \textbf{2.093 (0.577)} & \textbf{2.758 (3.697)} \\ \midrule

2D U-Net & 570 (100\%) & 0.937 (0.029) & 0.955 (0.016) & 0.948 (0.071) & 1.917 (0.294) & 1.888 (0.178) & 2.327 (2.566) \\
2D FCN & 570 (100\%) & 0.934 (0.032) & 0.958 (0.015) & 0.949 (0.078) & 1.913 (0.297) & 1.890 (0.347) & 2.161 (1.068) \\
3D U-Net & 570 (100\%) & 0.913 (0.112) & 0.945 (0.078) & 0.933 (0.093) & 2.104 (1.24) & 1.957 (0.68) & 2.722 (3.57) \\
MV U-Net & 570 (100\%) & \textbf{0.938 (0.027)} & \textbf{0.958 (0.013)} & \textbf{0.952 (0.079)} & \textbf{1.903 (0.345)} & \textbf{1.874 (0.142)} & \textbf{2.146 (1.004)} \\ \bottomrule
\end{tabular}%
 \begin{tablenotes}
 \item Approx. \# of conv weights (million)
 \item 2D U-Net: 0.8
 \item 2D FCN: 1.0
 \item 3D U-Net: 2.5
 \item MV U-Net: 1.2 
 \end{tablenotes}
 \end{threeparttable}
 }
\end{table}

\section{Discussion and Conclusion}
In this work, we presented a shape-aware multi-view autoencoder, a neural network capable of learning anatomical shape priors from multiple standard views, as well as a multi-view U-Net, a modification of the original U-Net architecture that incorporates the learned shape priors to improve the robustness of cardiac segmentation. In contrast to existing works which treat long-axis CMR segmentation and short-axis CMR segmentation as two separate tasks \cite{Bai2018,VIGNEAULT201895}, our approach, to the best of our knowledge, is the first that exploits the spatial context from the long-axis images to guide the segmentation on the short-axis images. The reported experimental results show that the proposed segmentation method not only demonstrates superior segmentation accuracy over state-of-the-art 2D baseline methods \cite{Bai2018,U-Net}, but also outperforms a 3D U-Net~\cite{Cicek2016}. This improvement is particularly evident on the basal and apical slices in the low-data setting, as expected. When training data is limited, segmenting these challenging slices particularly benefits from the additional anatomical information extracted from the LA views and injected into the segmentation network. Of note, our approach does not require a dedicated acquisition protocol, since LA images are routinely acquired in most CMR imaging schemes. Moreover, the proposed MV U-Net maintains the computational advantage of a 2D network, using fewer parameters ($\sim1.2$ million weights) than the 3D U-Net ($\sim 2.5$ million weights) during training. This advantage also contributes to the data efficiency of our method, achieving high segmentation performance with limited training data. Importantly, our method could be extended in the future to multi-structure cardiac segmentation. The proposed approach could also be potentially adopted to other medical image segmentation tasks.

\subsubsection{Acknowledgements.} 
This work was supported by the SmartHeart EPSRC Programme Grant (EP/P001009/1). Steffen Petersen acknowledges support from the National Institute for Health Research Barts Biomedical Research Centre. The cardiac multi-view image dataset has been provided under UK Biobank Access Application 18545.
\bibliographystyle{unsrt}
\bibliography{paper586}
\newpage
\begin{appendix}
\centering
\large
\textbf{Supplementary Materials}

\begin{figure*}[!htb]
    \centering
    \subfloat[Trained with 10\% data]{
    \includegraphics[width=0.5\textwidth]{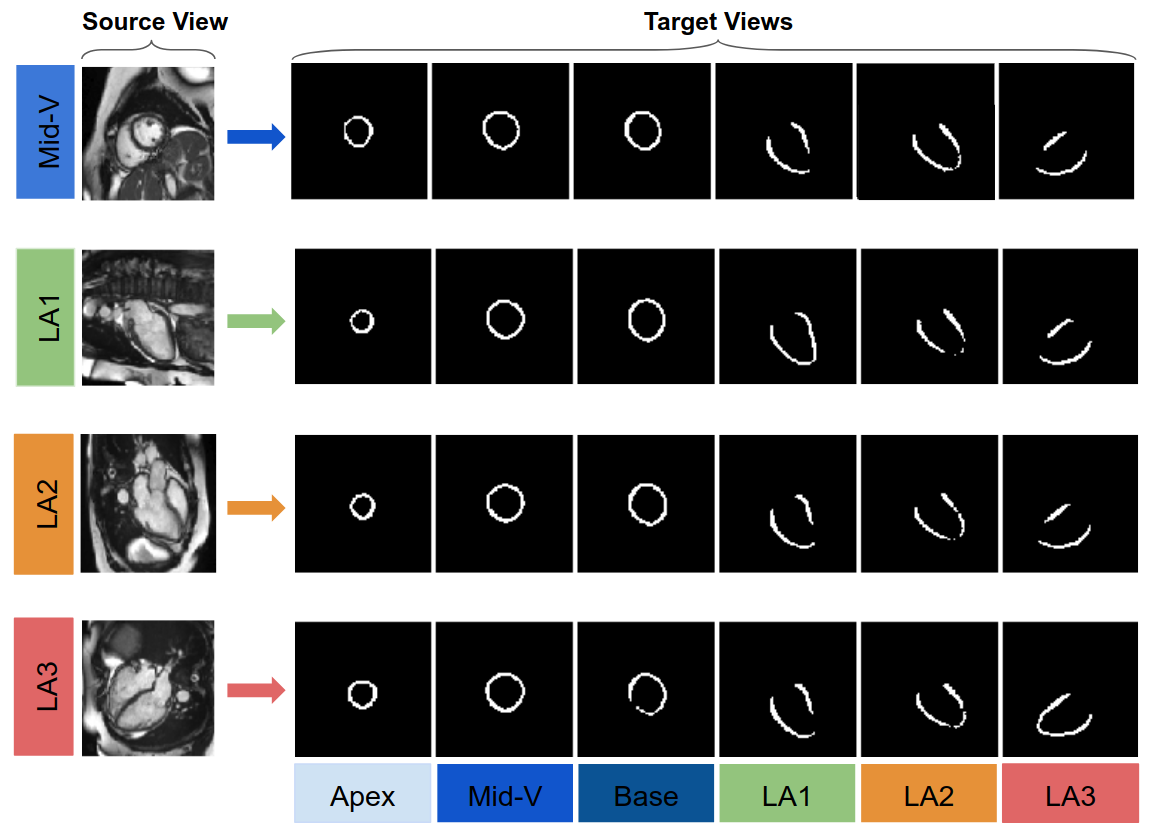}}
     \subfloat[Trained with 10\% data]{
    \includegraphics[ width=0.5\textwidth]{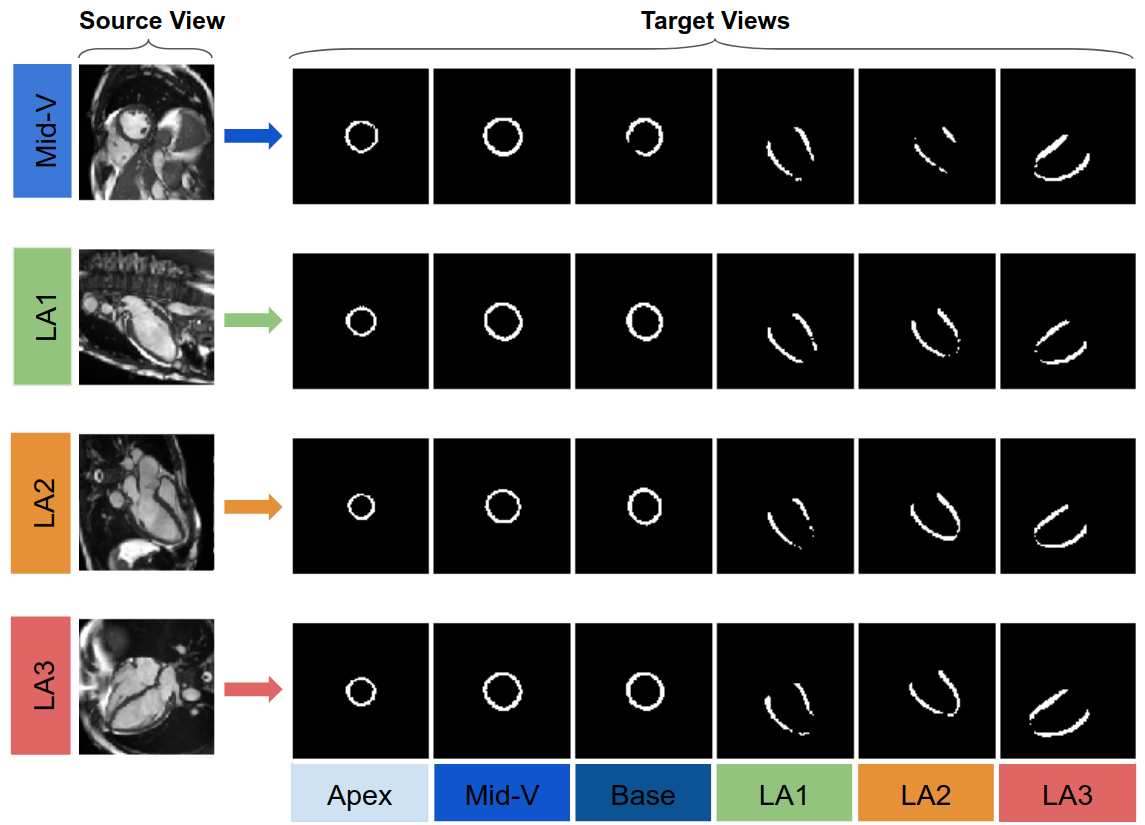}}
    \\
    \subfloat[Trained with 100\% data]{
    \includegraphics[width=0.5\textwidth]{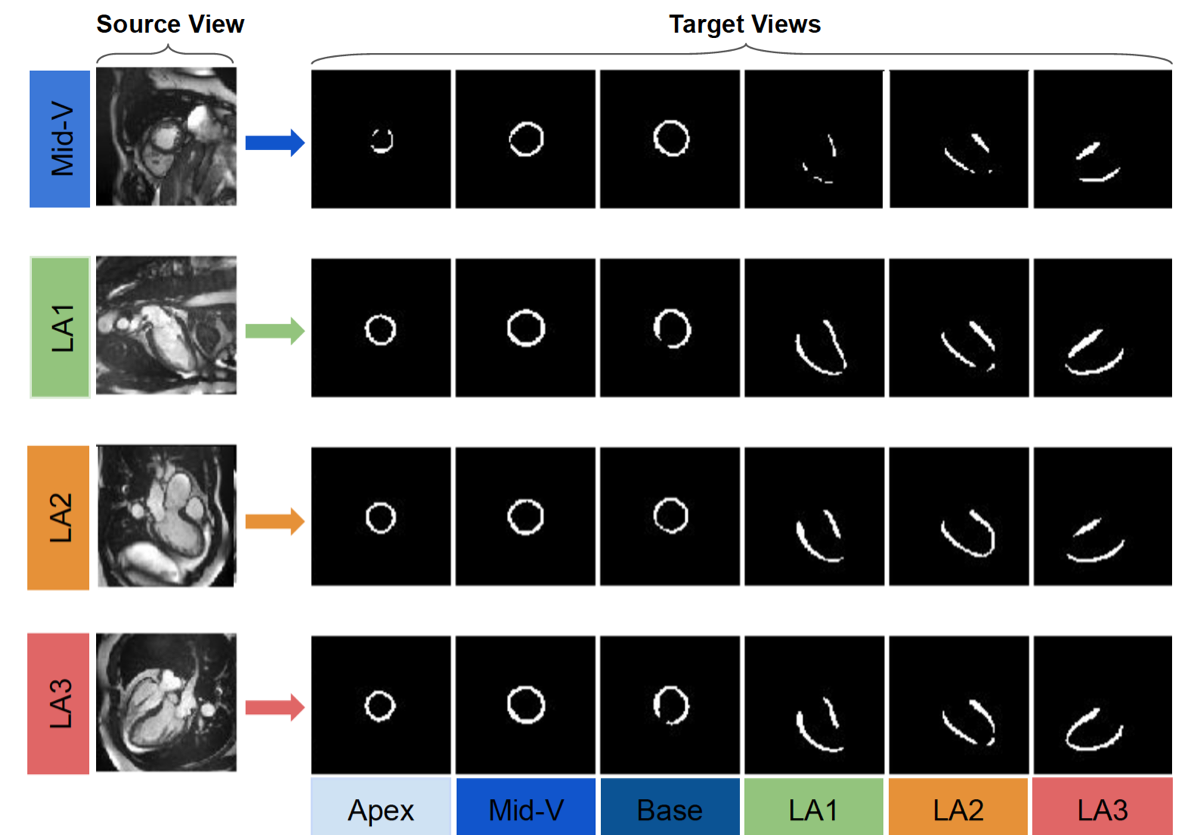}}
     \subfloat[Trained with 100\% data]{
    \includegraphics[ width=0.5\textwidth]{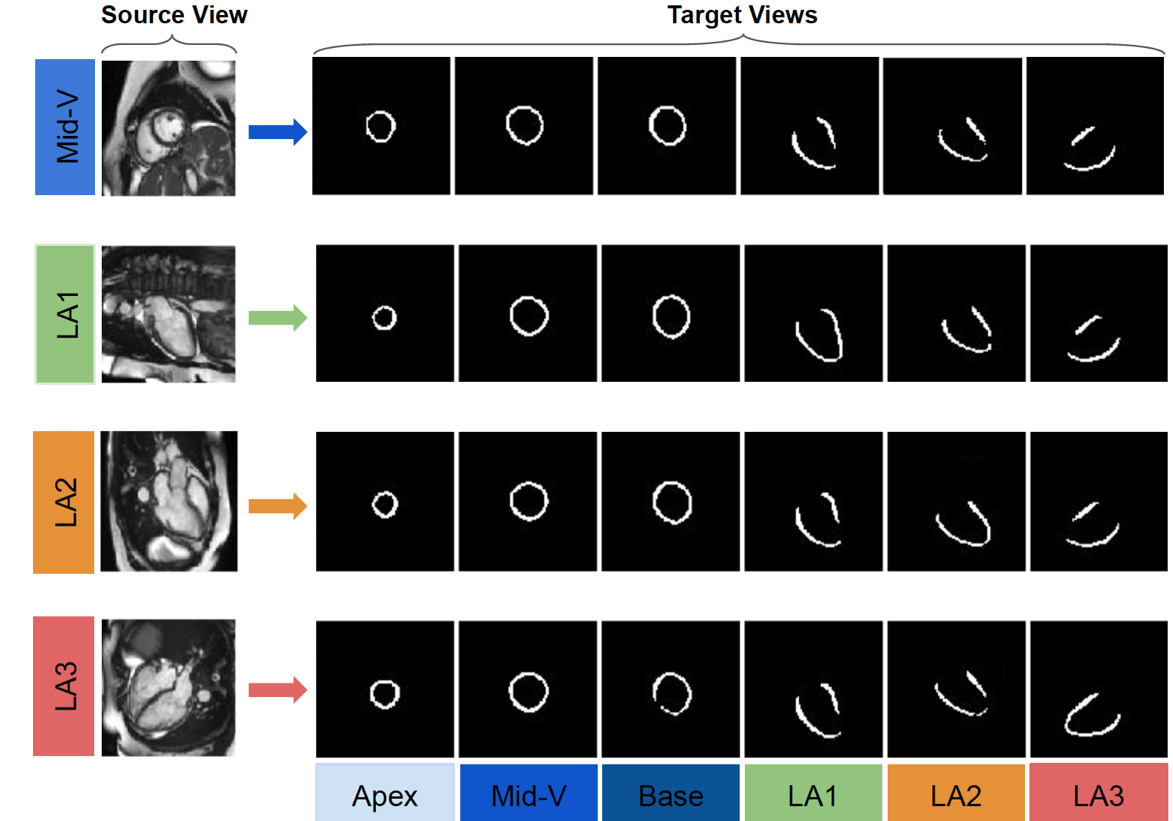}}
    \caption{Exemplar results of the proposed shape-aware multi-view autoencoder (Shape MAE) trained with 10\% and 100\% training data, respectively. Given only \textbf{one} source view (the first column) as input, the proposed Shape MAE is able to predict the myocardium shapes on the \textbf{six} target views (column 2 to column 7). This indicates that the proposed approach has the potential to encode the global shape characteristics of the myocardium in the latent space instead of a local embedding for a particular view of a subject.}
    \label{fig:shapeMAE_results}
\end{figure*}

\newpage
\begin{figure}[!htb]
    \centering
    \includegraphics[height=0.6\textheight, width=0.9\textwidth]{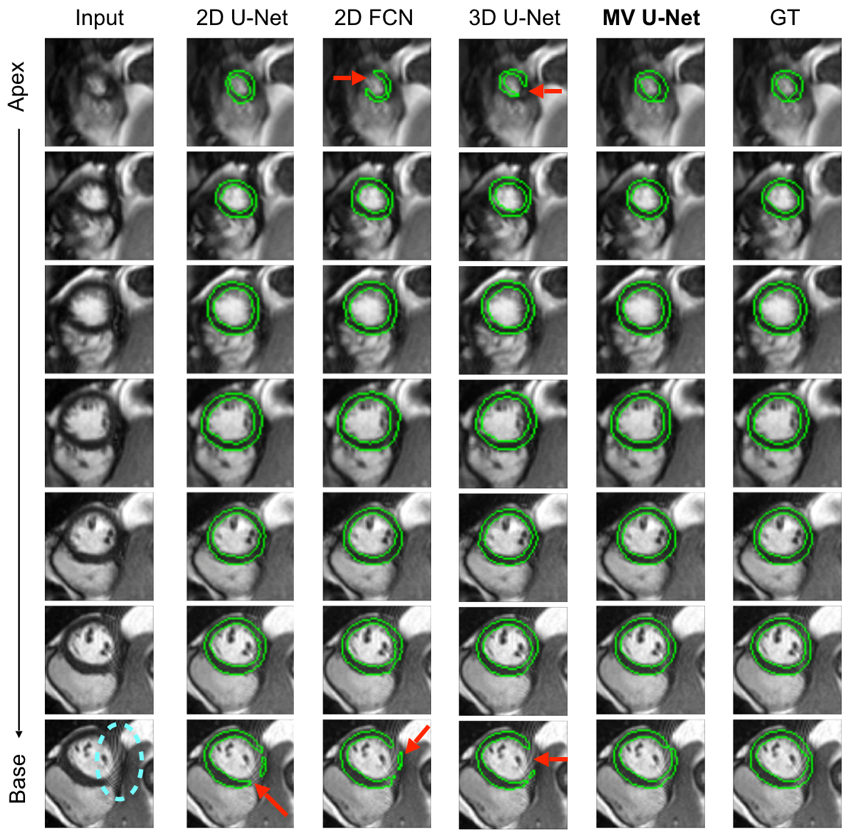}
    \caption{Example results of the proposed segmentation method (MV U-Net) and the baseline models (all trained with \textbf{100\%} training data) together with the ground truth (GT) on a stack of short-axis slices. Representative improvements for cardiac image segmentation can be observed when using the proposed method. For example, in contrast to the baseline models which produce poor results when there are unexpected artefacts on the image (see the region inside the \textcolor{cyan} {cyan ellipse}), the proposed method is able to properly identify the correct contours.}
    \label{fig:MV_UNET_results}
\end{figure}
\end{appendix}

\end{document}